\documentclass[sigplan,twocolumn,nonacm]{acmart}

\renewcommand\footnotetextcopyrightpermission[1]{}
\usepackage[normalem]{ulem}
\usepackage{xspace}
\usepackage{booktabs}
\usepackage{array}
\usepackage{multirow}
\usepackage{pifont}
\usepackage{graphicx}
\usepackage{enumitem}
\usepackage[ruled,vlined,linesnumbered]{algorithm2e}
\usepackage{tikz}
\usetikzlibrary{arrows.meta,fit}
\usetikzlibrary{arrows.meta,calc,decorations.pathreplacing,positioning,shapes.symbols}
\usepackage{pgfplots}
\usepgfplotslibrary{groupplots}
\pgfplotsset{compat=1.18}

\newif\ifnotes
\notesfalse

\newcommand{\sysname}{SAKURA\xspace}

\title{\sysname: Serving Attention Key-Value State Under Rate Adaptation
       for Private Image Editing on Phones}

\author{Shuncheng Jie}
\affiliation{%
  \institution{University of Waterloo}
  \country{Canada}}
\email{kevin.jie@uwaterloo.ca}

\author{Desen Sun}
\affiliation{%
  \institution{University of Waterloo}
  \country{Canada}}
\email{desen.sun@uwaterloo.ca}

\author{Ri Hong}
\affiliation{%
  \institution{University of Waterloo}
  \country{Canada}}
\email{r7hong@uwaterloo.ca}

\author{Yuyang Tian}
\affiliation{%
  \institution{University of Waterloo}
  \country{Canada}}
\email{y254tian@uwaterloo.ca}

\author{Hong Zhang}
\affiliation{%
  \institution{University of Waterloo}
  \country{Canada}}
\email{hongzhangblaze@gmail.com}

\author{Sihang Liu}
\affiliation{%
  \institution{University of Waterloo}
  \country{Canada}}
\email{sihangliu@uwaterloo.ca}

\date{}

\begin{document}
\pagestyle{plain}

\begin{abstract}

Diffusion transformers are widely used in image-editing services. 
To preserve both efficiency and privacy, existing systems commonly rely on cloud--device collaboration. 
Keeping request-specific inference on the device avoids sending personal images and prompts to the cloud. 
These workloads often use common templates, such as in face swapping and virtual try-on, enabling KV state to be precomputed and reused across requests.
However, this reusable state is large and scales with the number of templates, making it impractical to store all of them on mobile devices. 
Storing them in the cloud addresses this capacity issue, but cache transfer is placed onto the inference path and can even be slower than on-device computation, especially under fluctuating mobile networks.

We introduce \sysname, a cloud--edge architecture that delivers reusable attention state to mobile devices under a bandwidth budget.
It first uses a fine-grained token-importance-aware compression method to preserve image quality under limited transfer budgets.
At runtime, \sysname adaptively selects the cache representation for each upcoming shard based on network bandwidth.
To enable efficient execution, we develop a pipelined runtime that overlaps cache transfer and reconstruction with on-device execution.
We evaluate \sysname using three editing tasks and two diffusion models, using a Snapdragon 8 Elite Gen 5 phone.
\sysname reduces latency by 1.82$\times$ and 3.27$\times$ over CacheGen adapted to DiT and on-device generation, respectively.
It also reduces cache storage by 3.06$\times$ over uncompressed cache, while preserving image quality.

\end{abstract}

\maketitle

\section{Introduction}
\label{sec:introduction}

Diffusion transformers (DiTs) have become a leading architecture for
high-fidelity image generation and editing
\cite{peebles2023dit,esser2024sd3,chen2024pixart,
blackforestlabs2026flux2}. A widely deployed class of editing
applications personalizes only part of a provider-supplied image
while preserving the surrounding content. Virtual try-on \cite{kim2024stableviton, wan2026mobile,Zhao_2026_CVPR}, face swapping
\cite{10943471,yu2025magicmask,he2026highfidelitydiffusionfaceswapping}, and localized image editing
\cite{jiang2026flashps,MAG-Edit,simsar2024limelocalizedimageediting} are representative examples.
Personalized editing is already offered by cloud services,
such as Google Shopping's photo-conditioned virtual try-on and Amazon Nova
Canvas's managed virtual try-on API
\cite{google2025virtualtryon,amazon2025novavirtualtryon,
alibaba2026taobaovirtualtryon,alibabacloud2026characterswap}. These applications take
two kinds of different inputs: provider-owned templates can be
public and reused across users, whereas each request contains
privacy-sensitive user data such as personal photographs, reference images,
and editing instructions. Keeping request-specific inference on the device is
therefore attractive because these inputs and the intermediate states derived
from them need not be sent to a remote execution service. Recent advances in model compression, architecture optimization, and
heterogeneous mobile execution have made on-device diffusion inference
increasingly practical
\cite{li2023snapfusion,zhao2024mobilediffusion,chen2025heteroinfer}.

When requests share a template and a fixed mask defining the
editable region, computation associated with the unchanged region can be
reused across requests. In a DiT,
the key/value (KV) state of these reusable tokens can
be precomputed for each template
and reused while the device computes only
the request-dependent state for the edited region
\cite{jiang2026flashps,ijcai2026p150,fisedit,qin2025spotedit}. Cross-request reuse can therefore substantially
reduce repeated on-device computation. Every template has a different reusable state and the aggregate footprint grows with the number of templates. 

This creates a mobile-specific scalability gap between \emph{execution capacity} and \emph{reuse capacity}: a device may have enough compute and working memory to run each edit, yet lack the storage to retain caches across a large template catalog.
In our FLUX.2-klein-4B deployment \cite{blackforestlabs2026flux2} at $512\times512$ resolution, the reusable FP16 KV state occupies up to 1.46~GiB for a single template. A catalog with hundreds of templates would therefore require hundreds of gigabytes of persistent storage, far beyond what is practical to replicate on individual mobile devices.
Cloud-only execution avoids this storage constraint but requires sending private user requests to the
server. 
We instead consider a different cloud--edge split: request-specific
inference remains on the device, while the cloud stores the much larger,
provider-owned reusable templates and supplies the required cache on demand. 
Unlike conventional cloud--edge inference, which
offloads all or part of a request to remote compute \cite{han2016mcdnn,kang2017neurosurgeon,yan2024hybridsd}, this scheme uses the cloud  for template storage. 
The network carries reusable template state,
while user inputs and request-specific activations remain on the device.

Moving reusable state to the cloud resolves the device-side capacity problem,
but places cache transfer on the inference path. At the 545~Mbps median
bandwidth of the 5G trace used in our evaluation~\cite{5g_trace}, transferring
the full FP16 cache takes approximately 23~s, even without device-side cache
reconstruction. In comparison, executing the complete model locally without
cache reuse takes about 13~s on our mobile platform. Naively fetching
the full cache is therefore slower than running the complete model locally.
Remote reuse must both reduce the transmitted bytes and hide cache transfer
behind local computation.

These requirements translate into two challenges. First, the cache must be
compressed aggressively enough to arrive before use while preserving editing
quality. At an approximately 10\% mask ratio, only 2.3~s of local computation
is available in total to hide cache transfer; even uniform 2-bit quantization
requires at least 650~Mbps, above the median 5G bandwidth. 
Moreover, reusable cache compression must be constructed offline, before
future edit requests are known. This is challenging because the importance of cached tokens depends on the request-specific query generated by the edited region. 
Therefore, compression must determine which tokens to preserve without knowing the future request.
Second, cache transfer must adapt as mobile bandwidth changes during an edit.
A representation chosen early in execution may become too large when bandwidth
drops or unnecessarily compressed when bandwidth improves, requiring transfer
decisions to be revised as later cache state is consumed.

\noindent\textbf{Token-importance-aware cache compression.}
Our first key idea is to make request-dependent token importance usable for offline reusable cache compression. 
Although the exact attention scores depend on the specific request, We observe that, for a fixed template and mask, the reusable tokens that matter most remain largely consistent across different edit requests at the same transformer block and denoising
step.
This consistency allows importance measured from representative requests offline (i.e., through profiling) to guide compression for future requests.
Accordingly, compression can preseve cached tokens that are consistently important with higher accuracy. 

\noindent\textbf{Bandwidth-adaptive streaming.}
Our second key idea is to make cache transfer
decisions at fine granularity based on the network conditions.
Each shard contains the reusable state for a group of consecutive transformer blocks at one denoising step. 
\sysname can adjust its transfer decision before each shard is needed. 
It combines recent network measurements with device execution progress to determine how much cache data can arrive in time, then selects the highest-quality available representation that fits this budget. 
If an earlier transfer finishes ahead of use, the resulting lead time is carried forward to later shards. 
This allows cache transfer to track bandwidth changes within an edit instead of turning them into execution
stalls.

We present \sysname, a cloud--edge cache streaming system
that realizes these ideas through complementary offline and online
decisions. Offline, \sysname uses transferable token importance to construct
an importance-weighted low-rank factorization, preserving high-impact tokens'
KV state more accurately under aggressive compression. The compressed cache can vary along two dimensions: how much low-dimensional
structure is retained and how precisely it is encoded. These choices are
organized into a nested representation, allowing multiple compression levels
to share the same offline-constructed cache and enabling runtime adaptation
without repeating compression.
Online, \sysname adaptively selects the representation of each upcoming cache shard based on current network conditions and device execution progress, ensuring that the shard can be transferred before it is needed for computation.
To enable efficient mobile execution, we develop a heterogeneous mobile runtime that pipelines cache
reception on the CPU, reconstruction on the GPU, and model execution on the NPU, overlapping the communication and reconstruction behind on-device computation. 

This paper makes the following contributions:
\begin{itemize}[noitemsep,leftmargin=*]
\item  To the best of our knowledge, we present the first cloud--edge system that enables DiT cache reuse for mobile image editing, where reusable template cache states are stored in the cloud and streamed to the device upon request. 

\item We design a token-importance-aware cache compression scheme that exploits the stability of importance of the cache state across requests. 
We also design a cache representation that supports fine-grained adaptation across precision levels without storing a separate version for each level.

\item We develop an adaptive runtime that adjusts cache transfer to bandwidth fluctuations and overlaps cache transfer and reconstruction with on-device diffusion computation.

\item We evaluate \sysname on three editing tasks and two diffusion models using both 5G traces and WiFi.
On a Snapdragon 8 Elite Gen 5 phone, \sysname reduces generation latency by 1.82$\times$ and 3.27$\times$ compared with CacheGen \cite{cachegen} adapted to DiT and on-device generation, respectively.
It also reduces cache storage by 3.06$\times$ compared with uncompressed caching while preserving image quality.

\end{itemize}

\section{Background and Motivation}
\label{sec:background}

\subsection{Cross-Request KV Reuse for Image Editing}
\label{sec:2.1}

Diffusion transformers (DiTs) generate images through multiple denoising steps, each executing the same stack of transformer blocks. Within each block, token-wise operators process tokens independently, while self-attention aggregates global context through Query (Q), Key (K), and Value (V) representations \cite{mixfusion, vaswani2023attentionneed}.

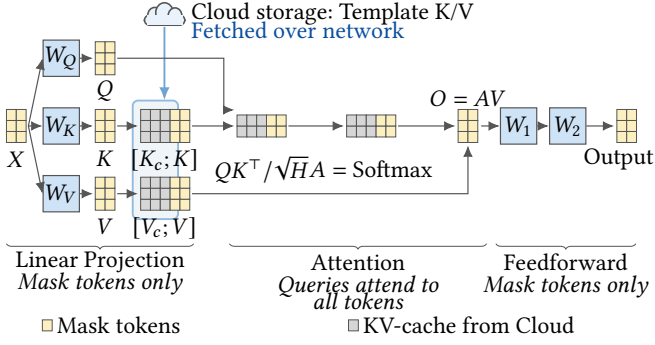
\begin{figure}[t]
\centering
\usetikzlibrary{arrows.meta,decorations.pathreplacing}

\definecolor{ccomp}{HTML}{FCEFC0}
\definecolor{ccache}{HTML}{D6D6D6}
\definecolor{cw}{HTML}{CFE3F7}
\definecolor{cline}{HTML}{555555}
\definecolor{ccloud}{HTML}{EAF2FB}
\definecolor{cnet}{HTML}{0B4F9E}
\definecolor{cnetlight}{HTML}{6FA3D2}

\begin{tikzpicture}[
  x=0.50cm,
  y=0.56cm,
  font=\small,
  arr/.style={
    -{Latex[length=1.5mm,width=1.3mm]},
    draw=cline,
    line width=0.45pt
  },
  netarr/.style={
    -{Latex[length=1.7mm,width=1.4mm]},
    draw=cnetlight,
    line width=0.7pt
  },
  weight/.style={
    draw=cline,
    fill=cw,
    line width=0.45pt,
    minimum width=4.8mm,
    minimum height=4.8mm,
    inner sep=0.5pt
  },
  txt/.style={
    inner sep=0pt,
    align=center
  }
]

\def\s{0.26}

\def\yQ{3.65}
\def\yK{2.05}
\def\yV{0.45}

\def\cQ{4.04}
\def\cK{2.44}
\def\cV{0.84}

\foreach \i in {0,1}{
  \foreach \j in {0,1,2}{
    \pgfmathsetmacro{\xx}{0+\i*\s}
    \pgfmathsetmacro{\yy}{\yK+\j*\s}

    \draw[
      fill=ccomp,
      draw=cline,
      line width=0.3pt
    ]
      (\xx,\yy) rectangle ++(\s,\s);
  }
}

\node[txt] at (0.26,1.66) {$X$};

\node[weight] (WQ) at (1.45,\cQ) {$W_Q$};
\node[weight] (WK) at (1.45,\cK) {$W_K$};
\node[weight] (WV) at (1.45,\cV) {$W_V$};

\draw[arr] (0.58,\cK) -- (WQ.west);
\draw[arr] (0.58,\cK) -- (WK.west);
\draw[arr] (0.58,\cK) -- (WV.west);

\foreach \basey/\lab/\laby in {
  \yQ/Q/3.27,
  \yK/K/1.67,
  \yV/V/0.07
}{
  \foreach \i in {0,1}{
    \foreach \j in {0,1,2}{
      \pgfmathsetmacro{\xx}{2.35+\i*\s}
      \pgfmathsetmacro{\yy}{\basey+\j*\s}

      \draw[
        fill=ccomp,
        draw=cline,
        line width=0.3pt
      ]
        (\xx,\yy) rectangle ++(\s,\s);
    }
  }

  \node[txt] at (2.61,\laby) {$\lab$};
}

\draw[arr] (WQ.east) -- (2.29,\cQ);
\draw[arr] (WK.east) -- (2.29,\cK);
\draw[arr] (WV.east) -- (2.29,\cV);

\filldraw[
  fill=cnetlight,
  fill opacity=0.10,
  draw=cnetlight,
  draw opacity=0.75,
  line width=0.7pt,
  rounded corners=2pt
]
  (3.34,0.24) rectangle (4.54,3.04);

\foreach \basey in {\yK,\yV}{

  \foreach \i in {0,1,2}{
    \foreach \j in {0,1,2}{
      \pgfmathsetmacro{\xx}{3.55+\i*\s}
      \pgfmathsetmacro{\yy}{\basey+\j*\s}

      \draw[
        fill=ccache,
        draw=cline,
        line width=0.3pt
      ]
        (\xx,\yy) rectangle ++(\s,\s);
    }
  }

  \foreach \i in {0,1}{
    \foreach \j in {0,1,2}{
      \pgfmathsetmacro{\xx}{4.33+\i*\s}
      \pgfmathsetmacro{\yy}{\basey+\j*\s}

      \draw[
        fill=ccomp,
        draw=cline,
        line width=0.3pt
      ]
        (\xx,\yy) rectangle ++(\s,\s);
    }
  }
}

\node[txt] at (4.18,1.64) {$[K_c;K]$};
\node[txt] at (4.18,0.04) {$[V_c;V]$};

\draw[arr] (2.93,\cK) -- (3.48,\cK);
\draw[arr] (2.93,\cV) -- (3.48,\cV);

\draw[netarr]
  (4.18,4.78) -- (4.18,3.05);

\draw[
  fill=ccloud,
  draw=cline,
  line width=0.6pt,
  rounded corners=2pt
]
  (3.65,4.82) rectangle (4.72,5.06);

\draw[
  fill=ccloud,
  draw=cline,
  line width=0.6pt
]
  (3.92,5.06) circle[radius=0.18];

\draw[
  fill=ccloud,
  draw=cline,
  line width=0.6pt
]
  (4.18,5.12) circle[radius=0.23];

\draw[
  fill=ccloud,
  draw=cline,
  line width=0.6pt
]
  (4.46,5.05) circle[radius=0.18];

\fill[ccloud]
  (3.72,4.84) rectangle (4.65,5.07);

\node[
  txt,
  anchor=west
]
  at (4.88,5.10)
  {Cloud storage: Template K/V};

\node[
  txt,
  anchor=west,
  text=cnet
]
  at (4.88,4.70)
  {Fetched over network};

\def\xAtt{6.10}

\foreach \i in {0,1,2}{
  \foreach \j in {0,1}{
    \pgfmathsetmacro{\xx}{\xAtt+\i*\s}
    \pgfmathsetmacro{\yy}{2.18+\j*\s}

    \draw[
      fill=ccache,
      draw=cline,
      line width=0.3pt
    ]
      (\xx,\yy) rectangle ++(\s,\s);
  }
}

\foreach \i in {0,1}{
  \foreach \j in {0,1}{
    \pgfmathsetmacro{\xx}{6.88+\i*\s}
    \pgfmathsetmacro{\yy}{2.18+\j*\s}

    \draw[
      fill=ccomp,
      draw=cline,
      line width=0.3pt
    ]
      (\xx,\yy) rectangle ++(\s,\s);
  }
}

\node[txt]
  at (6.75,1.42)
  {$QK^{\top}/\sqrt{H}$};

\draw[arr]
  (2.93,\cQ)
  --
  (5.75,\cQ)
  --
  (5.75,2.82)
  --
  (6.03,2.82);

\draw[arr]
  (4.91,\cK)
  --
  (6.03,\cK);

\def\xSoft{9.00}

\foreach \i in {0,1,2}{
  \foreach \j in {0,1}{
    \pgfmathsetmacro{\xx}{\xSoft+\i*\s}
    \pgfmathsetmacro{\yy}{2.18+\j*\s}

    \draw[
      fill=ccache,
      draw=cline,
      line width=0.3pt
    ]
      (\xx,\yy) rectangle ++(\s,\s);
  }
}

\foreach \i in {0,1}{
  \foreach \j in {0,1}{
    \pgfmathsetmacro{\xx}{9.78+\i*\s}
    \pgfmathsetmacro{\yy}{2.18+\j*\s}

    \draw[
      fill=ccomp,
      draw=cline,
      line width=0.3pt
    ]
      (\xx,\yy) rectangle ++(\s,\s);
  }
}

\node[txt]
  at (9.65,1.42)
  {$A=\mathrm{Softmax}$};

\draw[arr]
  (7.46,\cK)
  --
  (8.93,\cK);

\def\xOut{11.95}

\foreach \i in {0,1}{
  \foreach \j in {0,1,2}{
    \pgfmathsetmacro{\xx}{\xOut+\i*\s}
    \pgfmathsetmacro{\yy}{\yK+\j*\s}

    \draw[
      fill=ccomp,
      draw=cline,
      line width=0.3pt
    ]
      (\xx,\yy) rectangle ++(\s,\s);
  }
}

\node[txt]
  at (12.21,3.10)
  {$O=AV$};

\draw[arr]
  (10.36,\cK)
  --
  (11.88,\cK);

\draw[arr]
  (4.91,\cV)
  --
  (12.21,\cV)
  --
  (12.21,1.98);

\node[weight] (W1)
  at (13.55,\cK)
  {$W_1$};

\node[weight] (W2)
  at (14.85,\cK)
  {$W_2$};

\draw[arr]
  (12.56,\cK)
  --
  (W1.west);

\draw[arr]
  (W1.east)
  --
  (W2.west);

\foreach \i in {0,1}{
  \foreach \j in {0,1,2}{
    \pgfmathsetmacro{\xx}{16.15+\i*\s}
    \pgfmathsetmacro{\yy}{\yK+\j*\s}

    \draw[
      fill=ccomp,
      draw=cline,
      line width=0.3pt
    ]
      (\xx,\yy) rectangle ++(\s,\s);
  }
}

\draw[arr]
  (W2.east)
  --
  (16.08,\cK);

\node[txt]
  at (16.2,1.66)
  {Output};

\def\braceY{-0.34}

\draw[
  decorate,
  decoration={brace,mirror,amplitude=2pt},
  draw=cline,
  line width=0.35pt
]
  (0.00,\braceY)
  --
  (4.95,\braceY);

\node[txt]
  at (2.48,-0.78)
  {Linear Projection};

\node[txt]
  at (2.48,-1.28)
  {\emph{Mask tokens only}};

\draw[
  decorate,
  decoration={brace,mirror,amplitude=2pt},
  draw=cline,
  line width=0.35pt
]
  (5.90,\braceY)
  --
  (12.70,\braceY);

\node[txt]
  at (9.30,-0.78)
  {Attention};

\node[txt]
  at (9.30,-1.34)
  {\emph{Queries attend to}};

\node[txt]
  at (9.30,-1.70)
  {\emph{all tokens}};

\draw[
  decorate,
  decoration={brace,mirror,amplitude=2pt},
  draw=cline,
  line width=0.35pt
]
  (12.95,\braceY)
  --
  (16.65,\braceY);

\node[txt]
  at (14.80,-0.78)
  {Feedforward};

\node[txt]
  at (14.80,-1.34)
  {\emph{Mask tokens only}};

\draw[
  fill=ccomp,
  draw=cline,
  line width=0.3pt
]
  (0.95,-2.40) rectangle ++(0.26,0.26);

\node[
  txt,
  anchor=west
]
  at (1.33,-2.27)
  {Mask tokens};

\draw[
  fill=ccache,
  draw=cline,
  line width=0.3pt
]
  (9.05,-2.40) rectangle ++(0.26,0.26);

\node[
  txt,
  anchor=west
]
  at (9.43,-2.27)
  {KV-cache from Cloud};

\end{tikzpicture}
\caption{Cloud--edge KV reuse for template-based editing.}
\label{fig:attn-cloud-cache}
\end{figure}

Masked image editing modifies only a target region while leaving the rest of a
template unchanged, creating substantial cross-request reuse opportunities
\cite{jiang2026flashps,ijcai2026p150,fisedit,qin2025spotedit}. As illustrated in
Figure~\ref{fig:attn-cloud-cache}, token-wise operators only need to process
tokens inside the editable mask. For self-attention, the model computes Q, K,
and V for these masked-region tokens, combines their K and V with cached K and
V from the unchanged template tokens, and evaluates attention only for queries
from the masked region.
Such localized editing is common in applications including virtual try-on \cite{kim2024stableviton, wan2026mobile,Zhao_2026_CVPR}, face swapping \cite{swapmyface2026templates, canva2026faceswap}, and  inpainting \cite{xie2025turbofilladaptingfewsteptexttoimage,manukyan2024hdpainterhighresolutionpromptfaithfultextguided}.

This reuse enables sparse request-time execution, but requires a large reusable cache. Because the hidden states of unchanged tokens differ across denoising steps and DiT blocks, skipping their computation requires preserving the corresponding K and V tensors for \emph{every step and every block}. Thus, each template effectively maintains reusable KV state indexed by \emph{(denoising step, DiT block)}. For example, caching this state for a single $512\times512$ template in a lightweight 4B FLUX2 model~\cite{blackforestlabs2026flux2} requires 1.46~GiB, and the footprint grows with the template library.

Recent mobile diffusion systems demonstrate that modern mobile devices have
sufficient compute capability for on-device diffusion inference
\cite{feng2026dreamlitelightweightondeviceunified,li2023snapfusion,zhao2024mobilediffusion}.
However, this improvement in compute capability does not eliminate the storage
cost of maintaining a large library of per-template KV caches locally.
At the same time, keeping request-time inference local avoids transmitting privacy-sensitive user inputs and request-derived intermediate states to the cloud \cite{sun2021mind,nayan2024sok}. This creates a mismatch between \emph{compute capacity} and \emph{reuse-state capacity}: the device can execute the sparse request-specific computation, but cannot retain all reusable state needed to enable it. Supporting cross-request reuse at scale therefore requires separating request-time computation from reusable-state storage.

\subsection{Cloud--Edge Placement of Reusable State}
\label{sec:2.2}

The mismatch motivates a cloud--edge separation of computation and reusable state. As shown in Figure~\ref{fig:attn-cloud-cache}, the \emph{edge} performs all request-specific inference, while the \emph{cloud} stores the reusable per-template KV library and supplies the required entries on demand. This organization exploits the different roles of the two components: request-specific computation remains local, while reusable state, whose aggregate footprint grows with the template library, is placed in a remote tier with large capacity.

This setup differs from conventional cloud--edge inference, where the cloud primarily serves as an \emph{execution} resource. Prior systems either offload an entire request or partition model execution across the device and server \cite{han2016mcdnn,kang2017neurosurgeon}, with intermediate activations or generation stages crossing the network \cite{kang2017neurosurgeon,yan2024hybridsd,xie2025ecdiff,diffusionx2025}. In contrast, the cloud in our system acts as a remote \emph{storage for state}: computation does not migrate across the network; only reusable KV state is transferred to the device.

This separation solves the device-side storage problem, but shifts the bottleneck to cache transfer. On our mobile platform, full-model execution without reuse takes 13~s, whereas transferring the full FP16 cache over the 545~Mb/s median throughput of our 5G trace \cite{5g_trace} takes approximately 23~s. Naively fetching the full cache is therefore slower than local recomputation. Making remote reuse practical requires both reducing this transfer overhead and ensuring that the required cache state reaches the device in time for execution. We next examine the system challenges behind these requirements.

\section{Challenges and High-Level Ideas}
\label{sec:ideas}

In this section, we describe two key challenges in enabling streaming KV acache for on-device diffusion editing and then introduce our high-level ideas. 

\subsection{Quality-Efficient Cache Representation}
\label{sec:motiv-gap}

Due to limited mobile network bandwidth, only a limited cache size can be transferred before the reusable cache is consumed in the diffusion edit process.
As such, the cache compression must be \emph{quality-efficient}: for a given cache size, it should preserve the reusable information that contributes most to output quality. 
Using the same FLUX2 example in \S\ref{sec:background},  execution with cache reuse has a median aggregate transfer window of only 2.3~s at a 10\% mask ratio.
Even an aggressive 2-bit quantization still requires 650~Mbps of bandwidth, while leaving the quality non-ideal (results in \S\ref{sec:6.3}).
This motivates selectively preserving the cached information with high image quality impact under a fixed transfer budget.

\paragraph{Cross-request DiT reuse introduces a new token importance problem.}
Prior work has shown that token importance can reduce KV or computation.
LLM systems exploit this for request-time retention, eviction, or fetching~\cite{h2o,snapkv,infinigen}, while DiT serving systems use such signals for token routing, merging, pruning, or reuse within a denoising process~\cite{diffcr,importance_token_merge,topi}.
These approaches determine important tokens based on the ongoing request.

Cross-request DiT reuse is a new challenge: \emph{a reusable cache is constructed once for a template and must serve future unknown edit requests. }
Consequently, token-importance-aware compression cannot simply reuse important tokens measured
from the current request. Instead, it requires a template-level signal that remains
valid across future edit requests. 
This raises two questions: (1) whether some reusable cache tokens matter more than others to the final edit,
and (2) whether the same tokens remain important across different edit requests.

\begin{figure}[t]
\centering
\makebox[.48\columnwidth]{\includegraphics[width=0.52\linewidth]{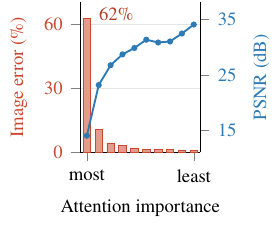}}
\hspace{0.01\columnwidth}
\makebox[.48\columnwidth]{\includegraphics[width=0.4\linewidth]{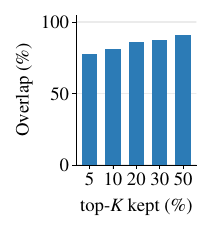}}

\makebox[.48\columnwidth]{(a) Uneven downstream impact}
\hspace{0.01\columnwidth}
\makebox[.48\columnwidth]{(b) Stable across requests}

\caption{Reusable template tokens have highly nonuniform downstream impact,
while their relative importance remains stable across request-specific
references.}
\label{fig:cache-pareto}
\end{figure}

\paragraph{Observation: token importance is uneven but stable across requests.}
We first measure how many reusable cache tokens affect the edited output. 
Across 50 templates,
we compute an importance score for every cached token at each transformer
block and denoising step using the squared attention received from
masked-region queries, 
following attention-aware KV distortion metrics~\cite{snapkv,pyramidkv,aatc}. We rank
cached rows by this score, divide them into ten equal groups, and
quantize one group each time to INT2 while leaving others exact.

Figure~\ref{fig:cache-pareto}(a) shows a highly skewed sensitivity
distribution: quantizing only the most-attended 10\% of cached tokens accounts
for 62\% of the induced image error, whereas perturbing lower-ranked groups
has substantially less effect. Thus, reusable cache tokens contribute very
unevenly to the final output quality.

For a fixed template, the edit mask and reusable template tokens are unchanged
across requests, while different edit requests produce different query vectors $(Q)$ for the masked region.
Figure~\ref{fig:cache-pareto}(b) shows high overlap across requests:
although the exact scores vary, the same reusable tokens tend to remain highly ranked. 
This consistency allows importance measured offline on
one set of requests to guide cache construction for future requests.

\paragraph{High-level idea.}
These observations suggest that reusable DiT KV caches should not be compressed
uniformly. Because token importance is both highly skewed and predictable
before future requests arrive, offline cache construction can use this
importance to preferentially preserve the information carried by high-impact
tokens while allowing larger approximation error on less influential ones.
The key is to incorporate token importance into the cache representation, 
while storing the cache representation in a way that can be efficiently transferred and reconstructed later online. \S\ref{sec:design-capture}
develops this idea using low-rank compression as an additional compression dimension beyond numerical precision.

\begin{figure}[t]
\centering
\includegraphics[width=\columnwidth]{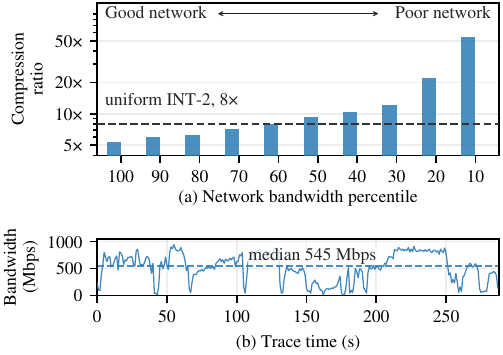}
\caption{Mobile network variation motivates adaptive cache streaming.
(a) Required compression ratio. (b) The underlying 5G bandwidth trace.}
\label{fig:bandwidth-compression}
\end{figure}

\subsection{Bandwidth-Adaptive DiT Cache Transfer}
\label{sec:motiv-perstep}

Even with an efficient cache representation, the cache transfer process still needs to be aware of mobile network bandwidth. 
Network conditions can change as a reusable DiT cache is consumed gradually throughout execution. 
Therefore, compression and transfer decisions must adapt to the granularity of DiT execution.

\paragraph{Mobile bandwidth changes within the DiT execution window.}
For requests with a 10\% mask ratio, on-device DiT execution takes 2.26~s.
To quantify the resulting transfer pressure, we examine the same duration windows
from a 5G trace~\cite{5g_trace} and compute the compression ratio
required to transfer the uncompressed FP16 cache within each window.
Figure~\ref{fig:bandwidth-compression}(a) shows that the median window requires
$9.3\times$ compression, while the slowest 10\% requires $53.8\times$.
Figure~\ref{fig:bandwidth-compression}(b) shows the corresponding throughput
variation over time.
Importantly, these variations occur on the same time scale as local DiT
execution. As a result, the bandwidth when an early cache shard is
transferred can differ substantially from that for later shards in
the same request. 
Consequently, cache compression and transfer must be adjustable throughout execution, rather than determined once. 

\paragraph{Token-level KV transfer does not match DiT cache consumption.}
Existing KV-transfer systems primarily target LLM context reuse or server-side
KV movement~\cite{cachegen,kvcodec,kvserve,lmcache}. For example, CacheGen~\cite{cachegen} partitions the cached context along the token dimension and
selects a transfer configuration for each token chunk. Each decision
covers the KV associated with that token chunk across multiple transformer
layers.
DiT consumes reusable KV differently. During an edit, the model advances
through transformer blocks and denoising steps, and the KV for each block is
needed only when execution reaches that block. 
Hence, token-level transfer granularity used by LLM systems does not align with DiT execution, as KV associated with the same token chunk is consumed at different transformer blocks and denoising steps.
For example, a cache shard needed by an early block must be transferred immediately, while one used by a later block can continue to adapt to subsequent bandwidth changes. 
Using the same token-level granularity couples cache data needed at different times, which can either over-compress some shards or delay others.
We instead use block-grained cache shards as the unit of transfer adaptation.
This aligns transfer decisions with cache consumption, allowing successive shards to exploit additional transfer time and adapt to subsequent bandwidth changes. 

\paragraph{High-level idea.}
Cache transfer adaptation should follow the consumption granularity of DiT execution.
Rather than fixing a transfer configuration for an entire request, the system can adjust cache transfer as successive shards are consumed by the on-device diffusion process.
This allows later shards to respond to changing mobile bandwidth while avoiding unnecessary quality degradation or transfer stalls.

\subsection{Summary of High-Level Ideas}

These challenges suggest two complementary mechanisms: construct a quality-efficient reusable cache offline, and adapt its transfer online as
network and execution conditions change. We next describe how \sysname
realizes these mechanisms in a cloud--edge system.
\section{Design}
\label{sec:design}

\begin{figure}[t]
\centering
\includegraphics[width=1\columnwidth]{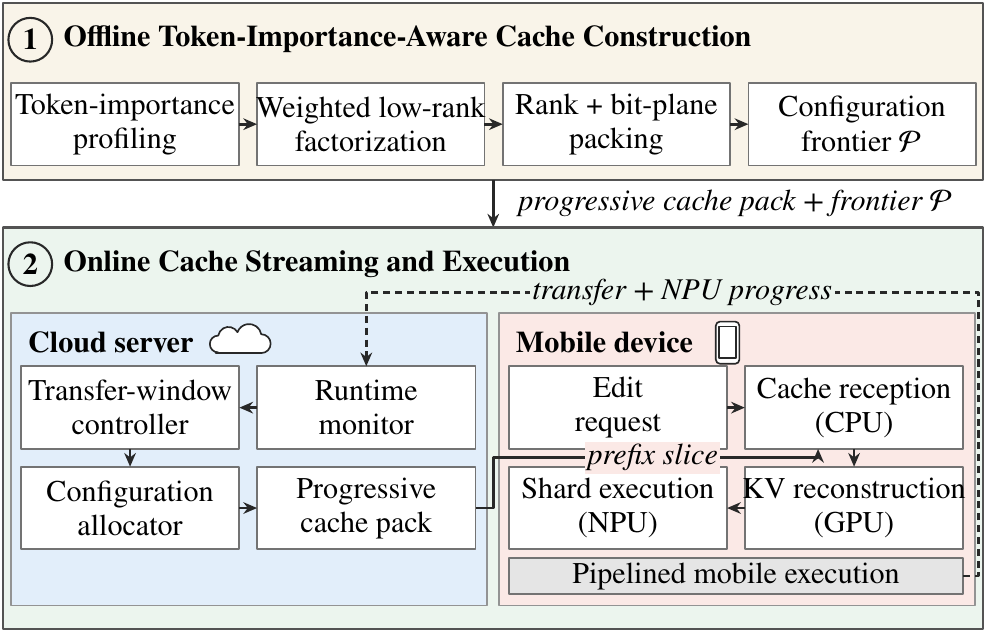}
\caption{\sysname system overview.}
\label{fig:system-overview}
\end{figure}

\subsection{Overview}

Figure~\ref{fig:system-overview} gives an overview of \sysname.
The design separates reusable cache construction from request-time transfer
and execution. \textbf{Offline}, \sysname builds two server-side artifacts for each template.
First, it uses profiled token importance to construct a compact low-rank KV
cache and stores its coefficient and basis matrices in a progressive layout,
so the same cache can serve multiple compression levels. Second, it profiles
a set of rank--precision trade-off points so later the runtime can select
the best configuration for image quality. \textbf{Online}, request-specific inference remains on the mobile device.
For each upcoming cache shard, the runtime monitor tracks recent network
bandwidth and device execution progress. The transfer-window controller uses
this information to determine how much cache can be transferred before the
shard is consumed, carrying forward unused transfer time from earlier shards.
The configuration allocator then selects the best-quality configuration from the
progressive cache pack and streams it to the device.

On the mobile device, cache reception, KV reconstruction, and shard execution
are pipelined across the CPU, GPU, and NPU, respectively, allowing cache
transfer and reconstruction to overlap model execution.

The remainder of this section describes each of these components.
\S\ref{sec:design-capture} presents offline token-importance-aware cache
construction and progressive packing;
\S\ref{sec:4.3} describes bandwidth-adaptive cache transfer;
and \S\ref{sec:design-pipeline} presents the heterogeneous mobile
execution pipeline.

\begin{figure*}[t]
\centering
\includegraphics[width=\linewidth]{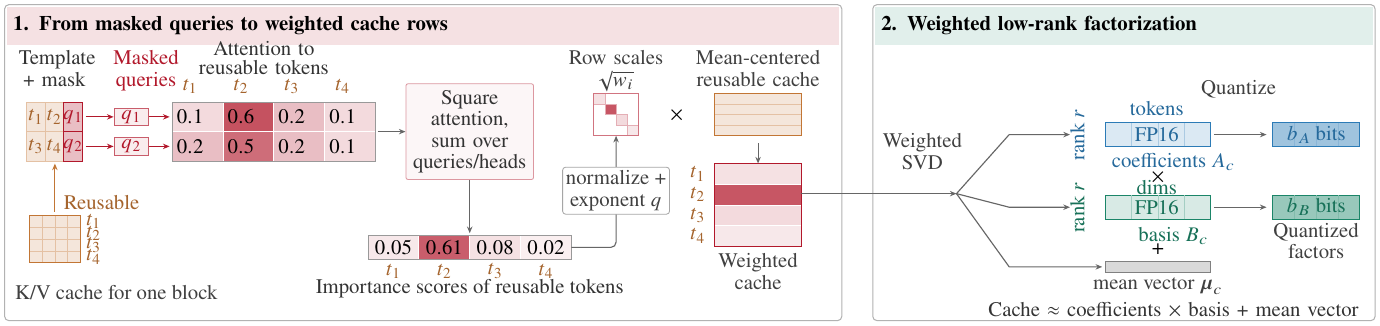}
\caption{Offline token-importance-aware cache construction. Masked-query attention is
aggregated into reusable-token importance scores and converted into
$q$-controlled row scales. These scales bias the low-rank factorization
toward important cache rows, after which the factors are quantized
independently.}
\label{fig:budget-fill}
\end{figure*}

\subsection{Offline Token-Importance-Aware Cache Construction}
\label{sec:design-capture}

Even aggressive quantization of the full KV tensor can exceed the transfer
budgets as shown in Figure~\ref{fig:bandwidth-compression}(a). We therefore first use low-rank
factorization to reduce the number of values that must be stored and
transmitted, and then quantize the resulting factors. 

For each transformer block and denoising step, we treat its reusable K or V
matrix $X_c\in\mathbb{R}^{N\times D}$ as one \emph{cache unit}, with one row
per reusable token. We compress each cache unit with a rank-$r$ approximation,
representing all token rows using $r$ shared components. Specifically, we use
truncated singular value decomposition (SVD)
~\cite{eckart1936approximation,golub1970singular}, which minimizes squared
reconstruction error for a given rank:
\begin{equation}
    X_c \approx A_c B_c + \mathbf{1}\mu_c^\top ,
    \label{eq:cache-low-rank}
\end{equation}
where $\mu_c$ is the row mean,
$A_c\in\mathbb{R}^{N\times r}$ contains the coefficients for each token, and
$B_c\in\mathbb{R}^{r\times D}$ contains the shared basis vectors.
When $r\ll D$, storing and transmitting $A_c$ and $B_c$ requires substantially
fewer values than the original $N\times D$ cache.

Low-rank factorization also makes the cache naturally progressive. SVD orders
components from larger to smaller contribution to cache reconstruction, so
transmitting additional components incrementally improves the approximation.
A smaller cache can therefore be obtained simply by stopping after fewer
components, without constructing a separate encoding for each rank.
Quantizing $A_c$ and $B_c$ provides a second way to vary the transmitted size.

Plain SVD, however, minimizes reconstruction error over all cache rows
uniformly. \S\ref{sec:motiv-gap} shows that reusable DiT tokens do not
contribute equally to the final edit, and that their relative importance can
be estimated offline. We therefore incorporate these importance scores when
constructing the low-rank factors. Rather than assigning a different rank to
each token---which would create variable-length coefficients and irregular
reconstruction---\sysname keeps one shared low-rank representation for the
entire cache unit and biases its construction toward accurately reconstructing
high-impact rows. This preserves a regular dense representation while making
the limited rank more useful for downstream editing.

\subsubsection{From masked queries to cache-row weights.}\label{sec:4.2.1}
The top half of Figure~\ref{fig:budget-fill} shows how \sysname turns the
importance signal from \S\ref{sec:motiv-gap} into construction-time
row weights. For each cache unit, \sysname measures how strongly masked-region
queries attend to each reusable template token. Following attention-aware KV
distortion metrics~\cite{snapkv,pyramidkv,aatc}, it squares and aggregates
attention across queries and heads, so tokens that consistently receive
stronger attention obtain larger importance scores. In
Figure~\ref{fig:budget-fill}, for example, the second reusable token receives
the largest score because both masked queries attend strongly to it.

\sysname normalizes the token-importance scores and uses an exponent $q$ to
control how strongly they influence the factorization. With $q=0$, all rows
are treated equally and the method reduces to ordinary SVD; increasing $q$
places progressively more emphasis on high-importance tokens. The resulting
weights are used only during offline construction and are not stored or
transmitted with the cache. We evaluate the effect of different $q$ values in \S\ref{sec:6.5.1}. 

\subsubsection{Weighted low-rank factorization.}
The bottom half of Figure~\ref{fig:budget-fill} shows how these weights affect
the low-rank representation. \sysname first subtracts the cache mean and
scales each row by the square root of its importance weight before applying
truncated SVD. This is equivalent to minimizing a weighted reconstruction
error in which errors on high-importance rows contribute more strongly
~\cite{srebro2003weighted,manton2003geometry}. The leading components
therefore devote more of the limited rank to preserving information carried
by important tokens.

After factorization, \sysname folds the inverse row scaling into the
coefficient matrix $A_c$. The resulting representation has the same regular
form as ordinary low-rank factorization: a coefficient matrix $A_c$, a shared
basis $B_c$, and a mean vector $\mu_c$. Thus, $q$ and the token weights affect
how the factors are constructed, but do not change the stored cache format.
The mobile device receives no importance weights and performs no
token-specific reconstruction; every cache unit is reconstructed using the
same dense matrix multiplication.

\subsubsection{Progressive rank and bit-plane packing.} \label{sec:4.2.3}
\begin{figure}[t]
\centering
\includegraphics[width=\columnwidth]{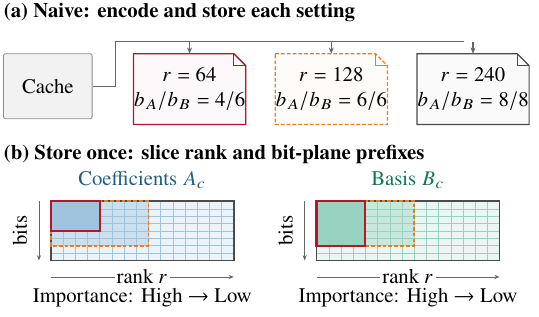}
\caption{Progressive cache packing. \sysname shares ordered rank components and precision bit planes across supported configurations.}
\label{fig:progressive-packing}
\end{figure}

Finally, as Figure~\ref{fig:budget-fill} shows, \sysname quantizes $A_c$ and
$B_c$ independently, using $b_A$ and $b_B$ bits. Their rank components remain
ordered by weighted singular contribution. This layout is what enables the
progressive representation in Figure~\ref{fig:progressive-packing}: instead
of storing a separately encoded cache for every $(r,b_A,b_B)$ configuration, \sysname stores the ordered factors once and later selects rank and
precision prefixes to match the available cache transfer budget.

\begin{figure}[t]
\centering
\includegraphics[width=\columnwidth]{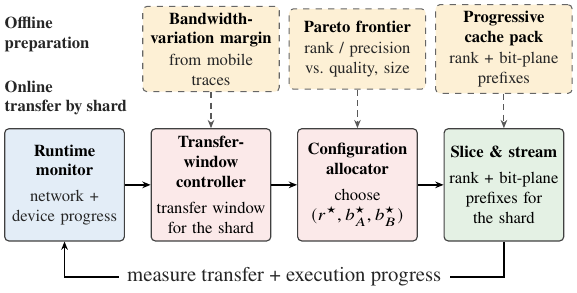}
\caption{Workflow of online adaptive cache streaming.
}
\label{fig:stream-flow}
\end{figure}

\subsection{Online Adaptive Cache Streaming}
\label{sec:4.3}

The offline phase leaves the server with the progressive cache pack described
above. At runtime, \sysname determines how much of this pack to transmit for
each cache shard as network and device conditions change.

\sysname operates at \emph{shard} granularity. A shard groups consecutive
transformer blocks executed together on the NPU, and the reusable KV state
consumed by one shard at one denoising step forms a cache shard. Because
shards are consumed sequentially, later decisions can incorporate
measurements from earlier transfers in the same edit.

Figure~\ref{fig:stream-flow} summarizes the online path. A runtime monitor
tracks network and device progress, a transfer-window controller determines
how long each upcoming shard can be transferred before use, and a
configuration allocator selects the corresponding prefix from the progressive
cache pack.

\subsubsection{Runtime monitor.}
The runtime monitor tracks the recent cache transfer rate and the arrival of
each cache shard. The device reports the bytes received and active receive
time, from which \sysname estimates the effective transfer rate $\hat{R}$. 
The monitor also records when each shard cache arrives relative to when it
is needed. If a shard arrives early, the remaining time is recorded as
carry-over $H_p$ and exposed to the transfer-window controller for the next
shard. Measurements are refreshed after every shard so that later decisions
reflect both recent network conditions and accumulated transfer lead.

\subsubsection{Transfer-window controller.}\label{sec:4.3.2}
\sysname adapts cache transfer to bandwidth: when bandwidth is high, it sends a
higher-quality cache representation, while under lower bandwidth it sends less
cache so that the transfer still completes before the shard is needed.
For an upcoming cache shard $p$, let $T_p^{\mathrm{ovlp}}$ denote the
remaining device-execution time over which its transfer can be hidden, and let
$H_{p-1}$ denote the carry-over time from the previous shard. The
controller sets the transmission window for shard $p$ to
$\gamma\left(T_p^{\mathrm{ovlp}} + H_{p-1}\right)$, where $\gamma<1$ provides
a safety margin against short-term bandwidth drops. Rather than predicting
future bandwidth, \sysname uses the recently observed transfer rate together
with this conservatively sized window.

\subsubsection{Configuration allocator.}\label{sec:4.3.3}
\begin{figure}[t]
\centering
\includegraphics[width=0.9\columnwidth]{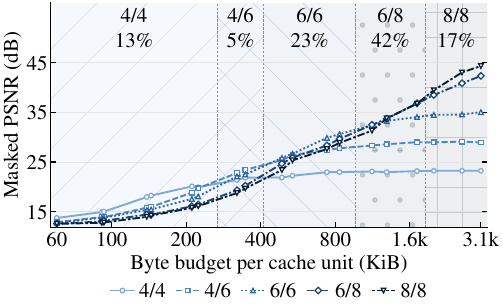}
\caption{Rank--precision tradeoff. Different
precision pairs are optimal at different byte budgets.}
\label{fig:bitwidth-monotune}
\end{figure}

A byte budget alone does not determine how it should be divided between rank
and numerical precision. Figure~\ref{fig:bitwidth-monotune} illustrates this
tradeoff. Under tight budgets, lower-precision factors can leave room for more
rank, whereas larger budgets increasingly favor higher precision. As a result,
no single rank--precision configuration dominates across all transfer budgets.

Offline, we therefore profile candidate combinations of rank $r$ and
bitwidths $(b_A,b_B)$ and retain only their nondominated configurations,
forming a small frontier $\mathcal{P}$. Given the current transfer rate and
transmission window, the allocator selects the configuration with the highest
profiled masked PSNR whose encoded shard fits:
\begin{align}
&(r^\star,b_A^\star,b_B^\star)
=
\arg\max_{(r,b_A,b_B)\in\mathcal{P}}
Q(r,b_A,b_B) \\
\text{s.t.}&
\quad
\frac{S_p(r,b_A,b_B)}{\hat{R}}
\le
T_p^{\mathrm{tx}}.
\label{eq:stream-selection}
\end{align}
$S_p(r,b_A,b_B)$ is the encoded size of shard $p$, and
$Q(r,b_A,b_B)$ is its masked PSNR measured during offline profiling.
Because $\mathcal{P}$ contains only nondominated configurations, the online
decision requires only a small enumeration.

Once a configuration is selected, the server directly slices the required
rank and bit-plane prefixes from the cache pack (Figure~\ref{fig:progressive-packing}(b)), without placing factorization
or re-encoding onto the request's critical path.

\begin{figure}[t]
\centering
\usetikzlibrary{arrows.meta,decorations.pathreplacing,positioning}

\definecolor{ccomp}{HTML}{FCEFC0}   %
\definecolor{ccache}{HTML}{E2E2E2}  %
\definecolor{cgpu}{HTML}{CFE3F7}    %
\definecolor{cline}{HTML}{555555}
\definecolor{cnet}{HTML}{1F6FB2}    %
\definecolor{cbad}{HTML}{C0392B}    %
\definecolor{cgood}{HTML}{1E7A44}   %

\providecommand{\bh}{0.32}
\providecommand{\lanebar}[5]{%
  \draw[fill=#4,draw=cline,line width=0.35pt,rounded corners=0.6pt]
    (#1,#3) rectangle (#2,{#3+\bh});
  \node[font=\footnotesize,inner sep=0pt] at ({0.5*(#1+#2)},{#3+0.5*\bh}) {#5};}

\begin{tikzpicture}[
  scale=1.45,
  lane/.style={font=\footnotesize,inner sep=1pt,anchor=east,align=right},
  pcap/.style={font=\footnotesize,inner sep=1pt},
  note/.style={font=\footnotesize,inner sep=1pt},
  hand/.style={-{Latex[length=1.1mm,width=0.9mm]},draw=cnet,line width=0.45pt},
  ax/.style={-{Latex[length=1.3mm,width=1.1mm]},draw=cline,line width=0.4pt},
]
\def\yC{0.88}\def\yG{0.44}\def\yN{0.00}

\begin{scope}[shift={(0,2.28)}]
  \node[lane] at (-0.10,{\yC+0.5*\bh}) {CPU};
  \node[lane] at (-0.10,{\yG+0.5*\bh}) {GPU};
  \node[lane] at (-0.10,{\yN+0.5*\bh}) {NPU};
  \lanebar{0.00}{1.10}{\yC}{ccache}{$j$}
  \lanebar{1.10}{1.55}{\yG}{cgpu}{$j$}
  \lanebar{1.55}{2.70}{\yN}{ccomp}{$j$}
  \lanebar{2.70}{3.40}{\yC}{ccache}{$j{+}1$}
  \lanebar{3.40}{3.85}{\yG}{cgpu}{$j{+}1$}
  \lanebar{3.85}{5.00}{\yN}{ccomp}{$j{+}1$}
  \draw[draw=cline!70,line width=0.4pt,dashed,rounded corners=0.6pt]
    (0.00,\yG) rectangle (1.10,{\yG+\bh});
  \node[note,text=cline] at (0.55,{\yG+0.5*\bh}) {start-up};
  \draw[draw=cline!70,line width=0.4pt,dashed,rounded corners=0.6pt]
    (0.00,\yN) rectangle (1.55,{\yN+\bh});
  \draw[draw=cbad,line width=0.45pt,dashed,rounded corners=0.6pt]
    (2.70,\yN) rectangle (3.85,{\yN+\bh});
  \node[note,text=cbad] at (3.28,{\yN+0.5*\bh}) {NPU stalls};
  \foreach \x in {1.10, 3.40}{
    \draw[hand] ({\x-0.20},\yC) -- (\x,{\yG+0.5*\bh});}
  \foreach \x in {1.55, 3.85}{
    \draw[hand] ({\x-0.20},\yG) -- (\x,{\yN+0.5*\bh});}
  \node[pcap,anchor=north] at (2.307,-0.10)
    {(a) Naive: no pipeline---prepare, then execute};
\end{scope}

\begin{scope}[shift={(0,0.42)}]
  \node[lane] at (-0.10,{\yC+0.5*\bh}) {CPU};
  \node[lane] at (-0.10,{\yG+0.5*\bh}) {GPU};
  \node[lane] at (-0.10,{\yN+0.5*\bh}) {NPU};
  \lanebar{0.00}{1.10}{\yC}{ccache}{$j$}
  \lanebar{1.10}{2.25}{\yC}{ccache}{$j{+}1$}
  \lanebar{1.10}{1.55}{\yG}{cgpu}{$j$}
  \lanebar{2.25}{2.70}{\yG}{cgpu}{$j{+}1$}
  \lanebar{1.55}{2.70}{\yN}{ccomp}{$j$}
  \lanebar{2.70}{3.85}{\yN}{ccomp}{$j{+}1$}
  \draw[draw=cline!70,line width=0.4pt,dashed,rounded corners=0.6pt]
    (0.00,\yG) rectangle (1.10,{\yG+\bh});
  \node[note,text=cline] at (0.55,{\yG+0.5*\bh}) {start-up};
  \draw[draw=cline!70,line width=0.4pt,dashed,rounded corners=0.6pt]
    (0.00,\yN) rectangle (1.55,{\yN+\bh});
  \foreach \x in {1.10, 2.25}{
    \draw[hand] ({\x-0.20},\yC) -- (\x,{\yG+0.5*\bh});}
  \foreach \a/\b in {1.55/1.55, 2.70/2.70}{
    \draw[hand] ({\a-0.20},\yG) -- (\b,{\yN+0.5*\bh});}
  \draw[draw=cline!70,line width=0.4pt,dotted] (3.85,{\yN-0.10}) -- (3.85,{\yN+\bh+0.10});
  \draw[draw=cline!70,line width=0.4pt,dotted] (5.00,{\yN-0.10}) -- (5.00,{\yN+\bh+0.10});
  \draw[{Latex[length=1.1mm,width=0.9mm]}-{Latex[length=1.1mm,width=0.9mm]},
        draw=cgood,line width=0.5pt]
    (3.87,{\yN+0.5*\bh}) -- (4.98,{\yN+0.5*\bh});
  \node[note,text=cgood] at (4.42,{\yN+\bh+0.16}) {saved vs.\ (a)};
  \draw[draw=cline,line width=0.35pt,dotted] (2.70,{\yN-0.02}) -- (2.70,{\yN-0.14});
  \node[note,anchor=north east] at (2.95,{\yN-0.11}) {release $j$};
  \node[pcap,anchor=north] at (2.307,-0.28)
    {(b) Pipelined: the three stages overlap};
\end{scope}

\draw[ax] (0,-0.28) -- (5.15,-0.28);
\node[note,anchor=south east] at (5.15,-0.27) {time};
\foreach \x/\c/\t in {-0.432/ccache/{CPU: network}, 1.318/cgpu/{GPU: reconstruct},
                      3.268/ccomp/{NPU: compute}}{
  \draw[fill=\c,draw=cline,line width=0.35pt] (\x,-0.76) rectangle ++(0.22,0.22);
  \node[note,anchor=west] at ({\x+0.28},-0.65) {\t};}
\end{tikzpicture}
\caption{Pipelined mobile execution. CPU reception and GPU cache
reconstruction for the next shard overlap NPU execution of the current shard.}
\label{fig:mobile-pipeline}
\end{figure}
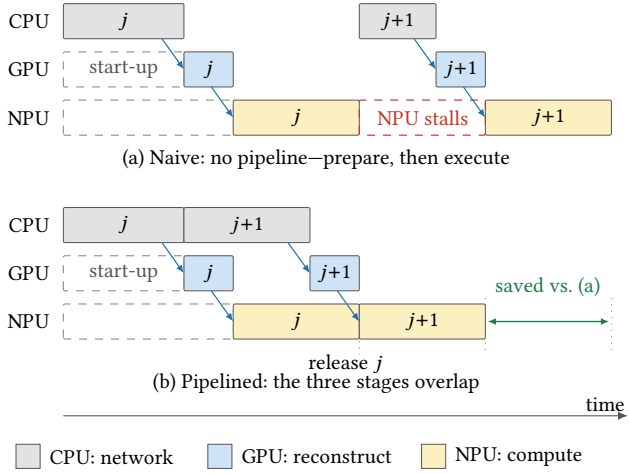

\subsection{Online Pipelined Mobile Execution}
\label{sec:design-pipeline}

As shown in Figure~\ref{fig:mobile-pipeline}, \sysname overlaps cache transfer,
reconstruction, and DiT execution across the CPU, GPU, and NPU.
After a short start-up phase, while the NPU executes shard $j$, the CPU
receives cache data for shard $j+1$ and the GPU reconstructs its KV tensors.
The reconstructed cache is buffered until the NPU reaches the next shard,
allowing the three stages to proceed concurrently instead of serializing on
each shard.
This pipeline hides most communication and reconstruction behind NPU
computation and avoids the NPU stalls incurred by the naïve
prepare-then-execute design.

\section{Implementation}
\label{sec:implementation}

\sysname's implementation focuses on Qualcomm platforms, which are widely used in smartphones. However, the design is generally applicable to other mobile platforms.

The server uses PyTorch and ExecuTorch~\cite{pytorch2024executorch} Qualcomm backend with an HTP v81 libso to compile the diffusion models for the mobile platform.
It first downloads the model from Huggingface then uses their diffusers pipelines, quantizes it to W8A16, and shards it by transformer block count so that each shard is one compiled graph; the
same block groups define the cache units in \S\ref{sec:4.3}.  Cache
capture follows FlashPS~\cite{jiang2026flashps}: we reuse its mask-aware
formulation of which KV state a template contributes, and store the result as
the per-unit packs in \S\ref{sec:4.2.3}.

The mobile device runtime implements the heterogeneous pipeline of
\S\ref{sec:design-pipeline} across all three engines in C++.
The CPU maintains the network transfer and the cache state machine: it requests the next shard cache, parses arriving
packets, and tracks which units are received, reconstructed, and released.  The
GPU dequantizes and reconstructs those units with OpenCL kernels.  The NPU
executes the compiled shards through the ExecuTorch
runtime~\cite{pytorch2024executorch}.  The three stages execute concurrently on separate shards, allowing each segment
to be received, reconstructed, and consumed in a pipeline without serializing
device execution on any single stage.

\section{Evaluation}
\label{sec:evaluation}

\subsection{Setup}
\label{sec:eval-setup}

\begin{figure}[t]
\centering
\includegraphics[width=.9\columnwidth]{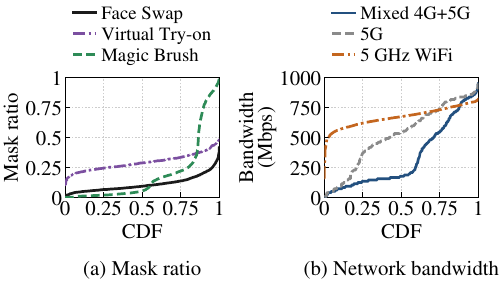}
\caption{Network and workload distribution.}
\label{fig:network-trace}
\end{figure}

\noindent\textbf{Tasks and datasets.}
We evaluate three image editing tasks. %
(1) The \emph{face swap} task replaces the human face in an image, where we take portrait templates from DiffusionDB~\cite{wangDiffusionDBLargescalePrompt2022} to cover diverse portrait styles and replacement faces from the Human-Art dataset~\cite{ju2023humanart}; the templates follow mask ratio distributions from Alibaba's Spring Festival dataset~\cite{jiang2026flashps}.
(2) The \emph{virtual try-on} task replaces clothes in a template image, where both images for clothes and templates are from the VITON-HD~\cite{choi2021viton} dataset. 
(3) The \emph{MagicBrush} task selects masked objects in an image and edits it with text, where images and text are from the
MagicBrush~\cite{Zhang2023MagicBrush} dataset.
Table~\ref{tab:workload} summarizes the dataset and Figure~\ref{fig:network-trace}(a) shows the mask ratio distributions. 

\noindent\textbf{Models and configurations.}
We evaluate the workloads above on two diffusion transformers that are suitable for mobile platforms:
FLUX2 (FLUX.2-klein-4B)~\cite{blackforestlabs2026flux2} with 4 steps and
SD3.5 (SD3.5-medium-turbo)~\cite{esser2024sd3} with 6 steps, both with default $512\times512$ resolutions.
Table~\ref{tab:workload} also lists the workloads, each consisting of a model and an image-editing task. FLUX2 is our primary model because it supports both image and text edit inputs for editing and can be applied to all three tasks; SD3.5 supports only text edit inputs and is evaluated only on the MagicBrush task.

\noindent\textbf{Template cache preparation.}
For each template, we profile it with 5 randomly chosen requests to identify the importance level of tokens which will be reused in evaluation. 
To ensure fair evaluation, we strictly separates data for profiling and evaluation.

\begin{table}[t]
\centering
\caption{Models and image editing tasks. }
\label{tab:workload}
\setlength{\tabcolsep}{2pt}
\begin{tabular}{ccccc}
\toprule
Workload & Task & Input Type & Model  & Mask \% \\
\midrule
\textbf{FS} & Face swap & Image+Image & FLUX2  & 11.2 \\
\textbf{VTON} & Virtual try-on & Image+Image  & FLUX2  & 26.3 \\
\textbf{MB} & \multirow{2}{*}{MagicBrush} & \multirow{2}{*}{Image+Text} & FLUX2 & \multirow{2}{*}{19.3} \\
\textbf{MB-SD} &  & & SD 3.5 & \\
\bottomrule
\end{tabular}
\end{table}

\noindent\textbf{Platform.}
All device measurements run on a OnePlus~15 phone with a Snapdragon 8 Elite Gen 5 SoC and 16 GB RAM. 
The phone is tested at room temperature of $24^\circ\text{C}$. 
The cloud is a server with a 12-core CPU and 64 GB RAM.
Template caches are stored on an SSD of the cloud server.

\noindent\textbf{Network traces.}
Our main experiments replay network bandwidth traces from the 5Gophers dataset~\cite{5g_trace}, which contains 603~seconds of per-second downlink throughput measurements collected by a handset on a commercial 5G network.
We shape and control the bandwidth with Linux \texttt{tc} on the host's interface~\cite{mao2017pensieve,akhtar2018oboe,fouladi2018salsify, cachegen}.
We also conduct an experiment using our institute's Wi-Fi 6 network, operating on the 5~GHz band, with the device placed at a stationary location.
This setting serves as an ideal scenario. 
CDFs of these traces are shown in Figure \ref{fig:network-trace}(b).
For each workload in Table \ref{tab:workload}, we randomly sample 100 starting points in the network trace and evaluate an edit request for each starting point.  

\noindent\textbf{Quality Metrics.}
We take the uncompressed cache (FP16) as the reference to compare against the design points.
\begin{itemize}[noitemsep,leftmargin=*]
\item \textbf{Peak Signal-to-Noise Ratio (PSNR)} compares each  design's image output with the reference pixel-wise. 
A higher value indicates better similarity, i.e., lower quality degradation after compression. 
\item \textbf{Structural Similarity Index Measure (SSIM)} shows the structural similarity between the two images. Higher value indicates better similarity.
\end{itemize}

\noindent\textbf{Baselines.}
We compare \sysname against uniform integer quantization (IntQ) and
CacheGen~\cite{cachegen}. 
IntQ uses one fixed bitwidth for the entire edit,
chosen from the initial bandwidth, with INT2 as the minimum precision. 
For CacheGen, we adapt the paper's open-source implementation to DiT KV tensors and mobile runtime. 

\subsection{Performance Evaluation}
\label{sec:eval-latency}

\begin{figure}[t]
\centering
\includegraphics[width=\columnwidth]{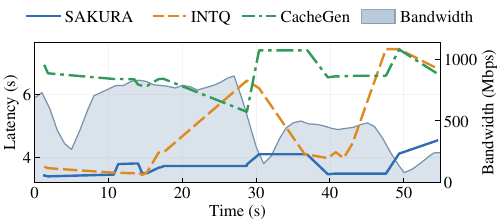}
\caption{An example showing FLUX2 latency under VTON workload during a 60-second network trace.}
\label{fig:demo60-vton}
\end{figure}

\noindent\textbf{Example trace.}
We first demonstrate the latency under a time-varying mobile network. 
Figure~\ref{fig:demo60-vton} shows FLUX2 + VTON serving under a 60-second 5G trace. 
The network bandwidth changes substantially over 
the trace, causing the fixed-compression baselines to exhibit large latency
variation. In particular, INTQ and CacheGen slow down
noticeably around at 30~s and 50~s, when their fixed cache representations
are too large for the available bandwidth.
In contrast, \sysname adapts the transmitted cache representation as bandwidth
changes, making latency stable. 

\begin{figure}[t]
\centering
  \includegraphics[width=\columnwidth]{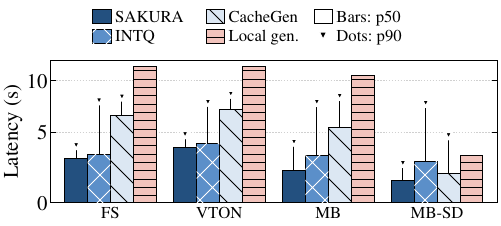}
  \caption{Generation latency under 5G trace.}
  \label{fig:5g-result}
\end{figure}

\begin{figure}[t]
\centering
  \includegraphics[width=\columnwidth]{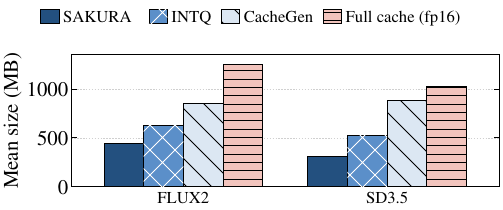}
  \caption{Per-template cache storage on server.}
  \label{fig:storage-overhead}
\end{figure}

\noindent\textbf{Generation latency.}
Figure~\ref{fig:5g-result} shows that \sysname achieves the lowest median and tail latency on all four workloads, reducing p90 latency by 44\% on average over INTQ and by 42\% on average over CacheGen.
INTQ remains competitive at the median, but its tail degrades sharply because its larger cache transfers make it more sensitive to bandwidth fluctuations.
CacheGen also exhibits substantially higher tail latency on the FLUX workloads because its coarse-grained transfer delays cache reconstruction and limits communication--computation overlap.
\sysname mitigates both issues through fine-grained, execution-aware cache transfer.
We also compare with local generation without cache reuse. The median latency of \sysname{} is 2.1--4.1$\times$ lower. 
Compared to FLUX2 workloads, MB-SD has a relatively lower speedup of 2.1$\times$ because SD3.5 is a lightweight model that has lower cost when running locally even without cache reuse. 

\noindent\textbf{Storage overhead.}
Figure~\ref{fig:storage-overhead} compares the per-template cache size, where \sysname has the lowest footprint for both models. %
The reduction comes from combining low-rank factorization with quantization, so a single compact representation supports multiple cache transfer budgets.
INTQ only reduces precision and has larger footprints.
CacheGen incurs the largest overhead, requiring 1.93$\times$ and 2.81$\times$ more storage than \sysname for FLUX2 and SD3.5, respectively, because each quality level stores an independent quantized cache with no sharing across levels.
With quality levels, this overhead grows nearly linearly, whereas \sysname reuses the same factorized representation across levels.

\subsection{Quality Evaluation}
\label{sec:6.3}

This section evaluates the quality compared to the reference scenario without compression (in FP16). 

\begin{figure}[t]
\centering
\includegraphics[width=\columnwidth]{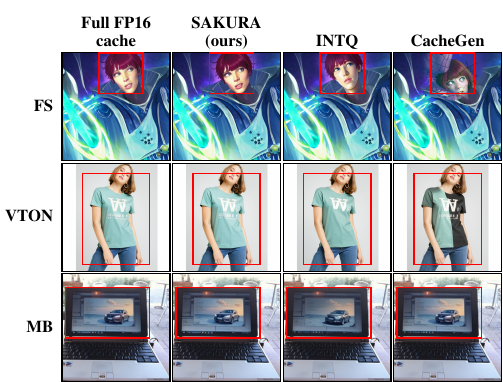}
\caption{Example outputs of requests from three tasks using FLUX2 with different systems.}
\label{fig:qualitative}
\end{figure}

\begin{figure*}[t]
    \centering
    \includegraphics[width=\textwidth]{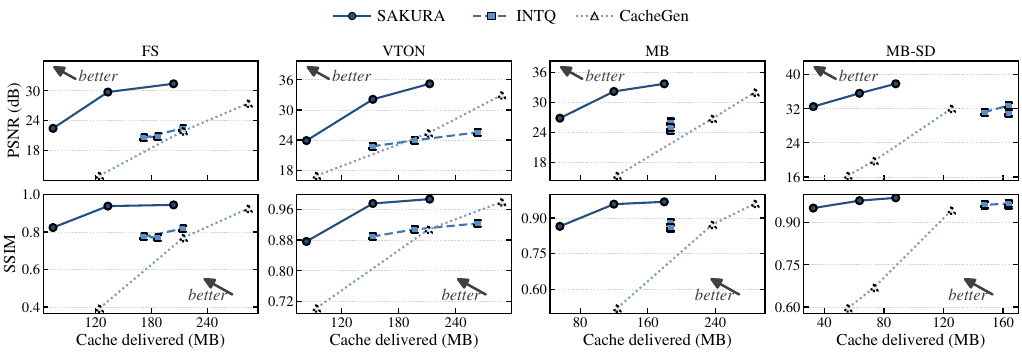}
    \caption{Transferred cache size vs. image quality.}
    \label{fig:pareto}
\end{figure*}

\noindent\textbf{Image Examples.}
Figure~\ref{fig:qualitative} shows representative outputs from face swapping, with a 10\% mask ratio and 400~Mbps bandwidth during editing. Compared with the uncompressed output, \sysname generates a mostly identical image. 
In comparison, INTQ has minor differences while CacheGen exhibits large deviations (especially in FS and VTON).

\noindent\textbf{Image quality.}
Table~\ref{tab:main_eval_quality_mean} shows that \sysname consistently achieves the highest mean PSNR and SSIM across all four workloads under 5G.
INTQ preserves more cache entries using uniform quantization, but does not allocate budget on the most critical components.
CacheGen degrades more severely because stronger compression evenly degrades cache components, without exploiting the static importance structure of DiT template caches.
Figure~\ref{fig:pareto} further illustrates this tradeoff: \sysname consistently forms the upper-left cache-size--quality frontier, achieving higher quality at the same transfer budget and requiring fewer bytes to reach a given quality level.
In particular, \sysname experiences low quality degradation as the cache budget decreases, whereas CacheGen's quality drops sharply.
Overall, \sysname transfers 30--56\% less data than INTQ and 23--42\% less than CacheGen while maintaining higher image quality.

\begin{table}[t]
\centering
\caption{Mean image quality on the real device (vs.\ full-cache output), averaged over all requests. }
\label{tab:main_eval_quality_mean}
\setlength{\tabcolsep}{2pt}
\begin{tabular}{llcccccc}
\toprule
 & & \multicolumn{2}{c}{5G (main)} & \multicolumn{2}{c}{4G+5G mix} & \multicolumn{2}{c}{WiFi} \\
\cmidrule(lr){3-4}\cmidrule(lr){5-6}\cmidrule(lr){7-8}
 & Systems & PSNR & SSIM & PSNR & SSIM & PSNR & SSIM \\
\midrule
\multirow{3}{*}{FS} & \sysname & \textbf{27.9} & \textbf{0.90} & \textbf{24.9} & \textbf{0.85} & \textbf{32.7} & \textbf{0.97} \\
 & INTQ & 21.3 & 0.79 & 21.0 & 0.78 & 21.4 & 0.80 \\
 & CacheGen & 20.8 & 0.70 & 16.2 & 0.51 & 23.5 & 0.81 \\
\midrule
\multirow{3}{*}{VTON} & \sysname & \textbf{30.4} & \textbf{0.95} & \textbf{26.8} & \textbf{0.91} & \textbf{37.3} & \textbf{0.99} \\
 & INTQ & 24.1 & 0.91 & 23.1 & 0.89 & 25.6 & 0.93 \\
 & CacheGen & 25.1 & 0.86 & 18.8 & 0.69 & 28.9 & 0.95 \\
\midrule
\multirow{3}{*}{MB} & \sysname & \textbf{30.9} & \textbf{0.93} & \textbf{27.3} & \textbf{0.86} & \textbf{34.2} & \textbf{0.969} \\
 & INTQ & 25.1 & 0.87 & 24.5 & 0.86 & 26.9 & 0.88 \\
 & CacheGen & 24.7 & 0.79 & 19.9 & 0.61 & 28.0 & 0.88 \\
\midrule
\multirow{3}{*}{MB-SD} & \sysname & \textbf{35.3} & \textbf{0.97} & \textbf{32.6} & 0.949 & \textbf{38.6} & \textbf{0.99} \\
 & INTQ & 31.6 & 0.96 & 31.3 & \textbf{0.96} & 32.0 & 0.96 \\
 & CacheGen & 22.7 & 0.74 & 19.0 & 0.66 & 23.4 & 0.75 \\
\bottomrule
\end{tabular}
\end{table}

\begin{table}[t]
\centering
\caption{User study comparing perceptual similarity to uncompressed cache.}
\label{tab:user-study}
\setlength{\tabcolsep}{6pt}
\begin{tabular}{lcccc}
\toprule
Method & FS & VTON & MB & MD-SD \\
\midrule
\sysname  & \textbf{86.0\%} & \textbf{88.0\%} & \textbf{90.0\%} & \textbf{85.1\%} \\
INTQ      & 56.5\% & 72.8\% & 79.7\% & 47.4\% \\
CacheGen  & 35.2\% & 55.8\% & 46.6\% & 43.5\% \\
\bottomrule
\end{tabular}
\end{table}

\noindent\textbf{User study.}
We further conducted a user study that involved 30 participants who were shown the uncompressed output as the reference and independently judged each baseline methods' outputs as \emph{similar} or \emph{not similar} to the reference. 
We first samples 30 requests out of randomizes the order of these outputs every time to avoid bias. 
Table~\ref{tab:user-study} shows that \sysname achieves the highest similarity score on every workload, ranging from 85.1\% to 90.0\%.
These user study results are consistent with the quality evaluation. 

\subsection{Sensitivity Studies}
\label{sec:eval-sensitivity}
We evaluate the primary model, FLUX2, for sensitivity. 

\begin{figure}[t]
\centering
\includegraphics[width=\columnwidth]{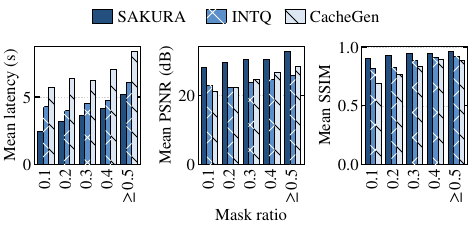}
\caption{Latency and quality under different editable-mask ratio.}
\label{fig:mask-ratio}
\end{figure}

\noindent\textbf{(1) Sensitivity to mask ratio.}
Figure~\ref{fig:mask-ratio} shows that larger masks increase
latency for all methods, as more tokens are involved in DiT computation.
At the same time, larger masks reduce the cache size to be transferred and increase the
per-step compute window, increasing the fraction of cache that can be transferred. 
As a result, compression becomes less aggressive and the quality gap between
systems narrows as the mask grows.
In contrast, smaller mask sizes combine the largest reusable cache with the shortest transfer window.
In such cases, \sysname provides the highest quality improvements over INTQ and
CacheGen as its token-importance-aware representation makes better use of the
transfer budget.

\begin{figure}[t]
\centering
\includegraphics[width=\columnwidth]{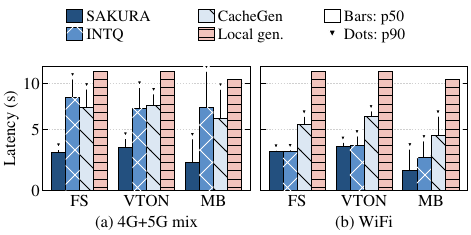}
\caption{Generation latency under different network traces.}
\label{fig:latency-sensitivity}
\end{figure}

\noindent\textbf{(2) Sensitivity to bandwidth.}
We evaluate two additional scenarios: a 4G+5G mixed trace that reflects lower network bandwidth and a WiFi scenario for a high-bandwidth connection. 
Figure~\ref{fig:latency-sensitivity} presents these scenarios. 
Under the 4G+5G mixed trace, \sysname reduces p90 latency by 55--76\% over INTQ and 47--62\% over CacheGen across the four workloads.
Under WiFi, where network is not a bottleneck, the gap narrows: on the three FLUX workloads, \sysname reduces p90 latency by up to 15\% over INTQ and 38--44\% over CacheGen.
These results show that \sysname provides the largest benefit when bandwidth is scarce, while its fine-grained transfer still benefits faster networks.

\begin{figure}[t]
\centering
\includegraphics[width=\columnwidth]{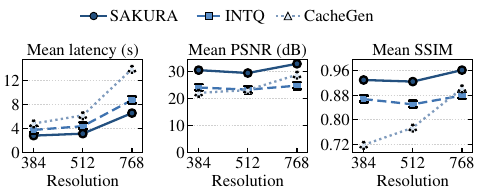}
\caption{Latency and quality under different resolutions. }
\label{fig:multi-res}
\end{figure}

\noindent\textbf{(3) Sensitivity to multi-image-resolutions.}
This experiment includes two other resolutions: $384\times384$ and $768\times768$, in addition to the default $512\times512$. 
Figure~\ref{fig:multi-res} shows that \sysname scales robustly across resolutions, consistently achieving the lowest latency and the highest PSNR/SSIM.
As resolution increases by 4$\times$ in pixel count, \sysname's latency and transferred cache grow by only 2.3--2.5$\times$, while ensuring image quality.
Overall, \sysname preserves its latency--quality advantage across all tested resolutions, demonstrating that its compression and streaming design scales effectively with larger DiT workloads.

\subsection{Ablation Studies}
\label{sec:eval-ablation}
This section studies the impact of three major mechanisms.

\begin{figure}[t]
\centering
\includegraphics[width=\columnwidth]{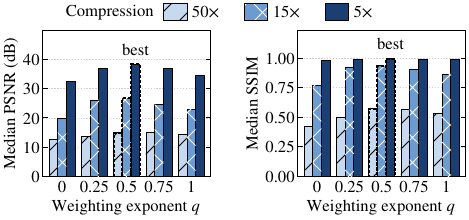}
\caption{Effect of attention-weighting exponent $q$.}
\label{fig:qsweep}
\end{figure}

\noindent\textbf{(1) Importance-weighting strength  (\S\ref{sec:4.2.1}).} \label{sec:6.5.1}
Figure~\ref{fig:qsweep} varies the exponent $q$ used for
importance weighting and shows results under different compression ratios: 5$\times$, 15$\times$, and 50$\times$. 
$q=0$ recovers plain SVD, while $q=0.5$ achieves the best overall quality, particularly under higher compression ratios.
We therefore use $q=0.5$ by default.

\begin{figure}[t]
\centering
\includegraphics[width=\columnwidth]{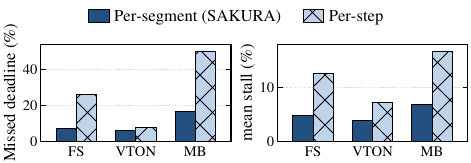}
\caption{Effect of per-shard adaptation.}
\label{fig:slo-ablation}
\end{figure}

\noindent\textbf{(2) Per-shard adaptation  (\S\ref{sec:4.3.2}).}
Figure~\ref{fig:slo-ablation} evaluates the benefit of revisiting transfer
decisions for successive cache shards. 
The left figure shows fine-grained adaptation reduces deadline violations (i.e., cache is not ready before the next block of diffusion computation starts) compared with a coarse transfer granularity.
We also note that this violation difference is low for VTON because it has the larger mask ratio and requires smaller caches to be transferred. 
The right figure shows the percentage of stall time due to missed deadlines, which follows a consistent trend. 

\begin{figure}[t]
\centering
\includegraphics[width=\columnwidth]{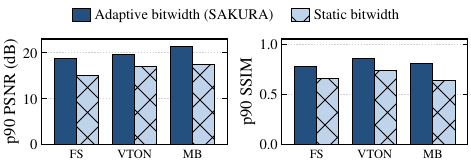}
\caption{Effect of adaptive bitwidth.}
\label{fig:bitwidth-ablation} 
\end{figure}

\noindent\textbf{(3) Adaptive bitwidth (\S\ref{sec:4.3.3}).}
Figure~\ref{fig:bitwidth-ablation} compares adaptive bitwidth with a fixed
8-bit configuration. Per-shard adaptation improves p90 PSNR by
2.5--3.9\,dB and p90 SSIM by 0.12--0.17 across the three workloads.
Under tight transfer budgets, using fewer bits per value leaves room to retain
more low-rank components. These results show that this precision--rank
trade-off is effective.

\section{Related Work}
\label{sec:related-work}

\noindent\textbf{Diffusion cache reuse.}
Prior work exploits substantial redundancy across diffusion denoising steps.
DeepCache and Faster Diffusion reuse intermediate U-Net features across nearby
steps, while Block Caching performs reuse at the granularity of individual
network blocks
\cite{ma2024deepcache,li2024fasterdiffusion,wimbauer2024blockcaching}.
For diffusion transformers, FORA caches attention and MLP outputs,
$\Delta$-DiT reuses block residuals, ToCa selectively reuses features for less
important tokens, and TeaCache adapts reuse intervals based on timestep-dependent
changes
\cite{selvaraju2024fora,chen2024deltadit,zou2025toca,liu2025teacache}.
These methods primarily reduce computation \emph{within} a single request by
exploiting temporal redundancy along the denoising trajectory.
Some prior work has considered reuse \emph{across} requests.
For example, NIRVANA reuses
intermediate noise states from similar prompts \cite{agarwal2024nirvana}, while FlashPS caches
template-dependent activations that can be shared across edits
\cite{jiang2026flashps}.
These systems assume server-side inference that directly reuses cached states on the server. 
In contrast, \sysname keeps request-specific processing on the mobile device while
streaming reusable template caches from the cloud.

\noindent\textbf{KV cache compression and memory optimization.}
KIVI and KVQuant compress KV caches through quantization
\cite{liu2024kivi,hooper2024kvquant}, while Palu stores low-dimensional
representations through factorized KV projections \cite{chang2025palu}.
StreamingLLM, H$_2$O, and SnapKV instead reduce the number of retained tokens
using recent-token, attention-importance, or observation-window-based
selection \cite{xiao2024streamingllm,h2o,snapkv}. InfiniGen selectively
prefetches important KV entries when caches are offloaded to host memory
\cite{infinigen}. For diffusion models, PTQ4DiT and ViDiT-Q reduce weight
and activation costs through timestep-aware quantization
\cite{wu2024ptq4dit,zhao2025viditq}. In contrast, we factorize
\emph{precomputed} template KV matrices offline and weight approximation
error using token importance shared across references of the same template.

\noindent\textbf{Cache transfer optimization.}
CacheGen compresses and streams reusable KV caches to LLM inference servers
with bandwidth-adaptive encoding \cite{cachegen}, while LMCache manages KV
movement across GPU, CPU, storage, and network tiers \cite{lmcache}.
KVServe selects compression profiles according to bandwidth, latency, and
quality constraints \cite{kvserve}, and KVCodec pipelines GPU-native
video-codec compression with KV transfer and reconstruction \cite{kvcodec}.
LLM.265 similarly applies hardware video codecs to tensor compression
\cite{xu2025llm265}, building on the broader idea of scalable,
quality-adaptive transfer \cite{schwarz2007svc}. Our setting differs in that
cache segments are streamed during on-device diffusion inference and must
arrive before the corresponding transformer blocks execute. Thus, transfer
decisions jointly depend on bandwidth and denoising execution progress.

\section{Conclusion}
\label{sec:conclusion}

In this work, we show that remote reuse of DiT template caches can make
on-device image editing practical without replicating large cache collections
on the device or offloading private request data to the cloud.
\sysname combines token-importance-aware offline compression with fine-grained,
bandwidth-adaptive cache streaming and a pipelined mobile runtime, allowing
cache transfer to track both network conditions and device execution.
Our evaluation demonstrates that \sysname simultaneously improves latency, storage efficiency, and output quality: it is 1.55--1.82$\times$ faster and 1.54--2.37$\times$ more storage-efficient than prior KV-streaming systems, while also outperforming fully local execution by 3.27$\times$ and reducing full-FP16 cache storage by 3.06$\times$.

\begingroup
\bibliographystyle{ACM-Reference-Format}
\bibliography{bib/references}
\endgroup

\end{document}